\documentclass[12pt]{article}
\usepackage{epsf,latexsym}
\usepackage{amsmath,amsfonts,amssymb,amsthm,cite}

\newtheorem{thm}{Theorem}[section]
\newtheorem{lem}[thm]{Lemma}

\newtheorem{pb}[thm]{Problem}

\theoremstyle{definition}                    
\newtheorem{defn}[thm]{Definition}
\theoremstyle{remark}
\newtheorem{rem}[thm]{Remark}

\numberwithin{equation}{section}             

\epsfverbosetrue
\newcommand{\double}[1]{\mathbb{#1}}

\newcommand{\rr}{\double{R}}
\newcommand{\zz}{\double{Z}}

\newcommand{\LL}{\double{L}}

\newcommand{\aaa}{\mathcal{A}}
\newcommand{\ccc}{\mathcal{C}}
\newcommand{\uu}{\mathcal{U}}

\newcommand{\dd}{\mathcal{D}}

\newcommand{\hh}{\mathcal{H}}

\newcommand{\jj}{\mathcal{J}}
\newcommand{\vv}{\mathcal{V}}

\newcommand{\ddd}{\,\hbox{$\partial\!\!\!/$}}

\newcommand{\de}{\hbox{\rm{d}}}

\newcommand{\pa}{\partial}

\newcommand{\bb}{\begin{eqnarray}}
\newcommand{\ee}{\end{eqnarray}}
\newcommand{\eee}{\nonumber\end{eqnarray}}

\newcommand{\qq}{\quad}

\begin{document}

\thispagestyle{empty}

\begin{center}

CENTRE DE PHYSIQUE TH\'EORIQUE $^1$ \\ CNRS--Luminy, Case 907\\ 13288
Marseille Cedex 9\\ FRANCE\\

\vspace{2cm}

{\Large\textbf{ Diffeomorphisms and orthonormal frames}}
\\

\vspace{1.5cm}

{\large Bruno Iochum $^2$, Thomas Sch\"ucker $^2$}

\vspace{1.5cm}

{\large\textbf{Abstract}}
\end{center}

There is a natural homomorphism of Lie pseudoalgebras from local vector
fields to local rotations on a Riemannian manifold. We address the question
whether this homomorphism is unique and give a positive answer in the
perturbative regime around the flat metric.
\vspace{1.5cm}

\vskip 1truecm

PACS-92: 11.15 Gauge field theories\\
\indent MSC-91: 81T13 Yang-Mills and other gauge theories

\vskip 1truecm

\noindent June 2004
\vskip 1truecm
\noindent CPT-2004/P.033\\
\noindent hep-th/0406213

\vspace{1,5cm}
\noindent $^1$ UMR 6207

- Unit\'e Mixte de Recherche du CNRS, des Universit\'es de Provence,  de la
M\'editerran\'ee et  du Sud Toulon-Var

- Laboratoire affili\'e \`a la FRUMAM - FR 2291\\
$^2$ Also at Universit\'e de Provence, \\ iochum@cpt.univ-mrs.fr,
schucker@cpt.univ-mrs.fr

\section{Introduction}

The spin lift $SO(3)\rightarrow SU(2)$ is an important part of quantum
mechanics. Indeed experiment teaches us that spinors, e.g. neutrons,
undergoing a rotation by $360^\circ$ differ from the non-rotated ones
\cite{neu},  and the action of the rotation group on spinors is double valued. In
special relativity this lift is extended to the Lorentz group,
 and in general relativity to the pseudogroup of coordinate  transformations.
As it stands, this latter extension is well defined only for genuine Riemannian
spin manifolds  (Euclidean signature) and goes in two steps: (i) from
coordinate transformations to local rotations of orthonormal frames, (ii) from
these rotations to spin transformations. The second step is double valued and
makes use of the spin structure.  We would like to know whether this general
relativistic lift is unique. By extension, such a lift will be called spin lift. 

In this note we ask the question on
infinitesimal level, which is much easier. Indeed for Lie algebras 
step (ii) is provided by the very definition of the spin cover and the issue
reduces to the uniqueness of the homomorphism from local vector fields to
infinitesimal rotations.
We prove this uniqueness  under the  assumption that the homomorphism
can be developed as a power series in
$h$ for a metric tensor 
$g=1+h$.

The spin lift from Riemannian geometry can be further extended
to noncommutative geometry
\cite{book,real,grav} where it allows to define the configuration space of
the spectral action \cite{cc} via the fluctuations of the Dirac operator. This use
of the spin lift motivated the question about its uniqueness. In a concluding
section, we summarize this
motivation together with an open problem that it raises.

\section{The pseudogroup of coordinate transformations}

Let $M$ be a smooth $n$--dimensional  Riemannian spin manifold with
metric
$g$. We are interested in the pseudogroup $\Gamma$ of (smooth) 
coordinate transformations between local charts of $M$ 
defining an atlas. This pseudogroup $\Gamma$ can be seen as 
the pseudogroup of local ($C^{\infty}$-) bidifferential transformations 
whose Jacobian, expressed in terms of $\Gamma_G$-coordinates, 
belongs to $G=GL(n,\rr)$. As a (locally flat) continuous pseudogroup, it possesses 
a Lie pseudoalgebra, which means that the presheaf over $M$ of 
sections of the sheaf of germs of differentiable vector fields 
($\Gamma_G$-vector fields over $M$) 
has a Lie algebra structure (see for instance \cite{spencer}).

Here, for simplicity, we prefer to write explicit coordinates in a chart; since we will have to manipulate
coordinates with possible singularities, globally defined diffeomorphisms are
not sufficient: let us choose an atlas
$(U_j,\alpha _j)$ of the manifold
$M$, where the
$U_j$ as well as all of their non-empty intersections are contractible open
subsets of $M$. The maps $\alpha _j$ are diffeomorphisms between the  $U_j$  
 and the open subsets $\alpha _j(U_j)$ of $\rr^n$. The pseudogroup $\Gamma$
 of
coordinate transformations  is a collection of the local diffeomorphisms of
$\rr^n$: 
$$ \sigma :\,\uu \longrightarrow \vv,$$  
$$\sigma:=\alpha _j\circ\alpha_i^{-1}: \qq \alpha _i(U_i\cap U_j) =:
\uu\longrightarrow
\alpha _j(U_i\cap U_j) =:\vv.$$

 Let $(\widetilde U_j,\widetilde \alpha _j)$ be an atlas of the orthonormal
frame bundle. We suppose again that the open subsets $\tilde U_j$  as well as all
of their non-empty intersections are contractible. We also suppose that their
projections on $M$ coincide with the $U_j$.

If our manifold is parallelizable, frames can be defined globally and we can
suppose that the
$\widetilde \alpha _j$ are of the form
$\big(\alpha _j, \ (T\alpha _j)^n\big)$. In general however,  $\widetilde
\alpha _j(\widetilde U_j)$ is a collection of frames defined only over $\alpha
_j(U)$. They are  orthonormal with respect to the metric induced on this open
subset of $\rr^n$ from the metric on $M$ by $\alpha _j$. As before, we define
the pseudogroup associated to this atlas whose open sets are the pre-images of
non-empty intersections $\widetilde \alpha _j\circ\widetilde \alpha_i^{-1}:
\widetilde \alpha _i(\widetilde U_i\cap \widetilde U_j)\longrightarrow
\widetilde
\alpha _j(\widetilde U_i \cap \widetilde U_j).$ Again we  simplify notations
and write such a diffeomorphism as
$$\widetilde \sigma :\widetilde \uu \rightarrow \widetilde \vv.$$ We denote a
generic element in $\widetilde \uu$ by $\big(x,e(x)\big)$, with
$x\in\uu\subset\rr^n$ and $e(x)$ an orthonormal basis, 
\bb e_a(x):= {(e^{-1})^\mu }_a(x)\,\frac{\pa}{\pa x^\mu }\,,\qq a=1,2,...,n.
\eee Orthonormality is with respect to the pull back of $g$ on $M$ to
$\uu$ by $\alpha ^{-1}_i$.

The metric tensor of this induced metric with respect to the
 coordinates $x^\mu $ is written 
\bb
 g_{\mu \nu }(x)=\left.g\left( \,T\alpha_i ^{-1}\frac{\pa}{\pa x^\mu }\,,
\,T\alpha_i ^{-1}\frac{\pa}{\pa x^\nu }\,\right)\right| _{\alpha _i(x)}\, .
\eee 
 
Following physicists' dangerous tradition, we use the same letter $e$ to denote
orthonormal frames and their coefficients with respect to holonomic frames
and we use the same letter $g$
 to denote the metric and its metric tensor. Then, orthonormality of $e$  is
equivalent to the matrix equation $ e^{-1\, T}\,g\,e^{-1}=1_n.$

We would like to write the transformation $\widetilde \sigma $ in these
notations. If
$M$ is not parallelizable we must include gauge transformations, i.e. local
rotations $\Lambda$:
\bb
\widetilde \sigma: =\big(\sigma ,(T_x\sigma )^n\circ\Lambda (x)\big),\qq
\Lambda :\uu\longrightarrow SO(n).
\eee Let us alleviate notations by dropping the tangent maps $(T_x\sigma )^n$
from the right hand side.  Then the composition law reads:
\bb (\tau  , \Omega )\circ (\sigma ,
\Lambda )= \big(\tau \circ \sigma , (\Omega\circ\sigma ) \Lambda \big),
\ee whenever target and source match. In terms of the frame bundle, this is
the cocycle condition. If targets and sources coincide we just have the
multiplication law in the semi-direct product, Diff$(\uu) \ltimes \,^\uu SO(n),$
where by $^\uu K$ we denote the group of smooth functions from $\uu$ to the group $K$.

\begin{defn}: A {\it pseudogroup homomorphism} between $\Gamma$ and the pseudogroup of
 local frame rotations is a collection of morphisms 
$L:\sigma \in {\rm Diff}(\uu,\vv) \mapsto \big(\sigma ,\Lambda (\sigma
,g)\big) \in {\rm Diff}(\uu,\vv) \ltimes \,^{\uu} SO(n)$, respecting
compositions whenever possible.

 The projection $p:(\sigma ,\Lambda )\mapsto \sigma $ is a homomorphism
from the pseudogroup of  gauge transformations to the pseudogroup of
coordinate transformations. 

{\it A lift} for $p$ is a map $L$ constructed from the metric only and such that $L
\circ p= $Id.
\end{defn} 
The dependence on the metric is indicated in $\Lambda (\sigma
,g)$ through the metric tensor $g$ on the source $\uu$ of $\sigma$. 

We denote by ${\jj_\sigma ( x) ^{\nu  }}_{\mu}:={\pa\sigma ^{\nu }(x) }/{\pa
x^\mu}$ the Jacobian of the coordinate transformation $\sigma$ and recall the
composition law $ \jj_{\tau \circ
\sigma}=\big(\jj_{\tau}
\circ
\sigma\big)
\jj_{\sigma}$.   Let us write $$\sigma \cdot g:=\left( \jj_\sigma
^{-1\,T}\,g\,\jj_\sigma ^{-1}\right)
\circ \sigma ^{-1}$$ for the push forward from $\uu$ to $\vv$ of the metric
tensor by
$\sigma
$ (pull back by $\sigma ^{-1}$). Then the composition law takes the form
$$(\tau \circ \sigma)\cdot g=\tau \cdot (\sigma \cdot g)$$
 and we have the
\begin{lem} A lift $L$ is a pseudogroup homomorphism if and only if
\bb 
\Lambda(1 ,g)=1_n,\qq\Lambda(\tau \circ\sigma ,g)=
\big(\Lambda(\tau ,\sigma \cdot g)\circ\sigma \big)\,\Lambda(\sigma ,g).
\label{comp}
\ee
\end{lem} A natural question is of course to construct a solution to
(\ref{comp}).

Here is a construction of such a lift based on a canonical way to orthonormalize
a basis $b_k$: Denote by $g_{k\ell}:= g(b_k,b_\ell)$ the positive matrix of
scalar products and define the orthonormal basis 
$e_a:={(g^{-1/2})^k}_a\,b_k$. 

Applying this procedure to the holonomic basis $\pa/\pa x^\mu $ over $\uu$,
we get the  orthonormal frame $e$ with coefficient matrix $e^{-1}={\sqrt
g}^{\,-1}.$ Following Raymond Stora we call this {\it the symmetric gauge}
because the coefficient matrix $e^{-1}$ is symmetric
\cite{bour,sch}. 

In the same manner we get an orthonormal frame $f$ over the target $\vv$
and the gauge transformations $\Lambda (\sigma ,g)$ is the unique rotation
that compares the orthonormal base $e$ and the pullback one $T\sigma ^{-1}f$,
namely 
$e_a=:{{\Lambda ^{-1}}^b}_a\, T\sigma ^{-1}f_b.$

Explicitly this gauge transformation is
\bb
\left.\Lambda(\sigma ,g)\right|_x =\left[\sqrt{\jj_\sigma ^{-1T}\,g\,\jj_\sigma
^{-1}}\,\jj_\sigma\, 
\sqrt{g^{-1}}\right]_x. 
\label{our}
\ee
 
This is the first step, from coordinate transformation to local rotations. The
second step to spin transformations is well known. The composition of the two
steps yields the desired  spin lift 
$\LL(\sigma ,g)=(\sigma ,S(
\Lambda(\sigma ,g))) $ where $S$ is the usual double valued homomorphism
\bb 
\left\{
\begin{array}{cl} S: SO(n) &\longrightarrow  \, Spin(n),  \\
\Lambda =\exp\omega  &\longmapsto \, \exp \left({\textstyle\frac{1}{8}} 
\omega_{ab} [\gamma ^{a},\gamma ^b]\right) 
\label{spin}
\end{array}\right.
\ee  with $\omega=-\omega ^T\,\in so(4)$. More explicitly, we get for any
spinor
$\psi$ and $x \in \uu=$ Source($\sigma)$
$$
\big(\sigma,\Lambda(\sigma,g)\big) \,
\psi\big(\sigma(x)\big)=S\big(\Lambda(\sigma,g)\big)(x)\, \psi(x).
$$ The spin structure on $M$ allows to glue $\LL$ together consistently.

If $M$ is  flat $\rr^3$ in Cartesian coordinates and $\sigma $ is a (rigid)
rotation
$R$,  then $\Lambda =R$ and $\LL$ is the spin lift of quantum mechanics.

A natural question arises: are there other lifts $\LL$, or an easier
question: are there  gauge transformations $\Lambda(\sigma ,g)$ other than
(\ref{our}) satisfying the multiplication law (\ref{comp})? 

We address the second question on infinitesimal level.

\section{The Lie pseudoalgebra of vector fields}

\subsection{Linearizing the composition law}

Consider the corresponding homomorphism
of Lie pseudoalgebras:
\bb \sigma (x)=x+\xi (x),\qq \tau (x)=x+\eta (x),\qq \Lambda(\sigma
,g)=1_n+\lambda (\xi ,g)+{\textstyle\frac{1}{2}} \lambda (\xi ,g)^2+...\ee where
$\xi $ and $\eta $ are vector fields on $\uu$ and the infinitesimal rotations
$\lambda \in so(n)$ are linear in the vector field. For infinitesimal coordinate
transformations, the composition law (\ref{comp}) reads
\bb
\lambda ([\xi ,\eta ],g)&=&[\lambda (\eta ,g),\lambda (\xi ,g)] -L_\eta \lambda
(\xi ,g)+L_\xi \lambda (\eta ,g)\cr \cr  &&-\,\frac{\delta }{\delta g} \lambda
(\xi ,g)\,\delta _\eta g +\,\frac{\delta }{\delta g} \lambda (\eta  ,g)\,\delta _\xi  g
\label{commut}
\ee  
where we write the functional derivative with respect to the metric tensor as 
\bb
\label{ict}
\frac{\delta }{\delta g} \lambda (\xi ,g)\,\delta g:=
\lambda (\xi ,g+\delta g)-\lambda (\xi ,g)+O(\delta g^2).
\ee
The variation of the metric tensor under an infinitesimal coordinate
transformation is denoted by
\bb
\delta _\xi g:=-\,\frac{\pa\xi }{\pa x}^T\, g-g\,\frac{\pa\xi }{\pa x}\,-L_\xi g,
\eee
the metric tensor $g$ being considered as matrix valued $0$--form.

\subsection{Linearizing the natural solution}

To linearize  equation (\ref{our}), the following lemma will be used repetitively.

\begin{lem}
Let X be a positive $n \times n$ matrix. Then, the solution of the constraint
\bb
\sqrt{g+X}=\sqrt{g} +{\textstyle\frac{1}{2}}
\sqrt{g}^{\,-1}I_g(X)+O(X^2)
\eee
is  
\bb I_g(X):= \int_{-\infty}^\infty\sqrt{g}^{\,\,it+1/2} \,X \, \sqrt{g}^{\,\,-it-1/2}
\,\frac{\de t}{\cosh(\pi t)}\,.
\ee
\end{lem}
\begin{proof}
We first remark that a solution of the equation $AY+YA=B$ where $A,B$ are
positive matrices with $A$ invertible is
\bb Y= \frac{1}{2}\, \int_{-\infty}^\infty A^{\,it-1/2} \,B \, A^{\,-it-1/2}
\,\frac{\de t}{\cosh(\pi t)}\,.
\eee 
as can be checked directly using complex integration (see for instance
\cite{Pedersen}). This solution is unique: if $Z=Z^*$ (as we may assume) is the
difference of two solutions, $AZ+ZA=0$, thus if $P_+$ is the projection on the
positive part $Z_+$ of $Z$, $P_+AZ_+ + Z_+AP_+=0$,
$A^{\frac{1}{2}}Z_+A^{\frac{1}{2}}=0$ since the spectrum of $P_+AZ_+$ and
$A^{\frac{1}{2}}Z_+A^{\frac{1}{2}}$ are equal (up to a possible zero) and
$Z_+=0$; similarly  $Z_-=0$.

Using analytical properties of the square root, we can write
$\sqrt{g+X}=\sqrt{g} +B+O(X^2)$  with $ \vert \vert B \vert\vert \leq
c\vert\vert X\vert \vert$. Thus 
$g+X=(\sqrt{g} +B)(\sqrt{g} +B)+O(X^2)=g+\sqrt{g}B+B\sqrt{g} + B^2 + O(X^2)$
and $X=\sqrt{g}B+B\sqrt{g}$, so $B=\frac{1}{2}\sqrt{g}^{\,-1}\, I_g(X)$
according to the previous remark.
\end{proof}

A few useful formulas are
\bb  I_{1_n}(X)=X,\qq I_{g^{-1}}(X)=\sqrt{g}^{\,-1}I_g(X)\sqrt{g},\qq
\int_{-\infty}^\infty t^{2\ell}
\,\frac{\de t}{\cosh(\pi t)}\,=\,\frac{E_\ell}{2^{2\ell}}\, ,
\eee with the Euler numbers $E_1=1,$ $E_2=5$, $E_3=61$,\,\dots

\noindent Now, the  linearized  equation (\ref{our}) can be written as
\bb
\lambda (\xi ,g)={\textstyle\frac{1}{2}} \sqrt{g}^{\,-1}I_g
\left( -\,\frac{\pa\xi }{\pa x}^T\,g-g\,\frac{\pa\xi }{\pa
x}\,\right)\sqrt{g}^{\,-1}+
\sqrt{g}\,\frac{\pa\xi }{\pa x}\,\sqrt{g}^{\,-1}.
\eee
However, strangely enough, both infinitesimal versions, that of the composition
law and that of the natural solution, are more involved than their finite
versions. In order  to continue nevertheless, we will resort to perturbation
theory around the flat metric. 

To this end, the functional derivative (\ref{ict}) of the natural solution $\lambda
$ with respect to the metric tensor is
\bb
\,\frac{\delta }{\delta g} \lambda (\xi ,g)\,\delta g&=& -{\textstyle\frac{1}{2}}
\sqrt{g}\,\frac{\pa\xi }{\pa x}\,g^{-1} I_g(\delta
g)\sqrt{g}^{\,-1}+{\textstyle\frac{1}{2}}\sqrt{g}^{\,-1}I_g(\delta g)
\,\frac{\pa\xi }{\pa x}\,\sqrt{g}^{\,-1}\cr \cr  &&+{\textstyle\frac{1}{2}}
\sqrt{g}^{\,-1}I_g(\delta _\xi ^-\delta g)\sqrt{g}^{\,-1} -{\textstyle\frac{1}{4}}
\sqrt{g}^{\,-1}I_g(\delta _\xi ^- g)g^{-1} I_g(\delta g) \sqrt{g}^{\,-1}\cr\cr  
&& -{\textstyle\frac{1}{8}} \sqrt{g}^{\,-1}I_g(\sqrt{g}^{\,-1}I_g\big(\delta g)
\sqrt{g}^{-1}I_g(\delta _\xi ^-g)\big)\sqrt{g}^{-1}\cr \cr &&
-{\textstyle\frac{1}{8}} \sqrt{g}^{-1}I_g(\sqrt{g}^{\,-1}I_g\big(\delta _\xi ^-g)
\sqrt{g}^{\,-1}I_g(\delta g)\big)\sqrt{g}^{\,-1},
\eee with the abbreviation
\bb
\delta _\xi^- g:=-\,\frac{\pa\xi }{\pa x}^T\, g-g\,\frac{\pa\xi }{\pa x}\,.
\eee 
Now we can compute its Taylor series around the unit metric
tensor
\bb 
\lambda (\xi ,1_n+h)&=& {\textstyle\frac{1}{2}} \left(
\,\frac{\pa\xi }{\pa x}\,-\,\frac{\pa\xi }{\pa x}^T\,\right)+
{\textstyle\frac{1}{8}}h\left( \,\frac{\pa\xi }{\pa x}\,+\,\frac{\pa\xi }{\pa
x}^T\,\right)  -{\textstyle\frac{1}{8}} \left( \,\frac{\pa\xi }{\pa
x}\,+\,\frac{\pa\xi }{\pa x}^T\,\right)h\cr \cr &&
-{\textstyle\frac{1}{16}}h^2\left( \,\frac{\pa\xi }{\pa x}\,+\,\frac{\pa\xi
}{\pa x}^T\,\right)  +{\textstyle\frac{1}{16}} \left( \,\frac{\pa\xi }{\pa
x}\,+\,\frac{\pa\xi }{\pa x}^T\,\right)h^2 +O(h^3).
\label{series}
\ee

 \subsection{Uniqueness of the Taylor series}

\begin{thm} \label{th}
Any Taylor series satisfying the infinitesimal composition law (\ref{commut})
is either identically zero or coincides with the Taylor series of the natural
solution (\ref{series}). 
\end{thm}

\begin{proof}

Let us view the commutator (\ref{commut}) as a first order functional
differential equation. We may hope to have a unique solution if we have an
initial condition at $g=1_n$. This initial condition $\lambda (\xi ,1_n)$ is an
antisymmetric $n\times n$ matrix, linear in $\xi $ and covariant under rigid
rotations of the Cartesian coordinates $x$:

We suppose that the solution
$\lambda (\xi ,g)$  can be written as a Taylor series in the components of $\xi $
and $g$ and of their derivatives. Then the initial condition, $g=1_n$, does not
contain derivatives of the metric tensor and it must be of the form
\bb
\lambda (\xi ,1_n)= a\left( \frac{\pa\xi }{\pa x}\,-\,\frac{\pa\xi }{\pa
x}^T\right) +a_1\left( \frac{\pa\Delta \xi }{\pa x}\,-\,\frac{\pa\Delta \xi }{\pa
x}^T\right) +a_2\left( \frac{\pa\Delta^2 \xi }{\pa x}\,-\,\frac{\pa\Delta^2 \xi
}{\pa x}^T\right)+\cdots
\eee where $\Delta $ is the (flat) Laplacian. The first derivative $\frac{\delta
}{\delta g} \lambda (\xi ,1_n)\,\delta g$ is an antisymmetric matrix, linear in
$\xi $ and $\delta g$:
\bb
\frac{\delta }{\delta g} \lambda (\xi ,1_n)\,\delta g&=& \left(b \,\frac{\pa\xi
}{\pa x}\,+c\,\frac{\pa\xi }{\pa x}^T\,\right)\delta g\ +\ 
\left(b_{11} \,\frac{\pa\Delta \xi }{\pa x}\,+c_{11}\,\frac{\pa\Delta \xi }{\pa
x}^T\,\right)\delta g\cr\cr &&  + \left(b _{12}\,\frac{\pa\xi }{\pa
x}\,+c_{12}\,\frac{\pa\xi }{\pa x}^T\,\right)\Delta \delta g\ +\ 
\left(b_{13} \,\frac{\pa\xi_\alpha  }{\pa x}\,+c_{13}\,\frac{\pa\xi_\alpha 
}{\pa x}^T\,\right)\delta g_\alpha
\cr \cr  &&- \textup{ transposed + higher derivatives}.
\eee 
Plugging these two series into equation (\ref{commut}) with $g=1_n$
leaves us with two solutions for the coefficients $a$, $b$, $c,\, \dots$ There is the
trivial solution where all coefficients vanish. The other solution is 
\bb a={\textstyle\frac{1}{2}} ,\qq b=c=-{\textstyle\frac{1}{8}},
\eee all higher coefficients vanish. We now have our initial condition
\bb \lambda (\xi ,1_n)= {\textstyle\frac{1}{2}} \left( \,\frac{\pa\xi }{\pa
x}\,-\,\frac{\pa\xi }{\pa x}^T\,\right)
\eee 
and equation (\ref{commut}) determines its first derivative,
\bb
\frac{\delta }{\delta g} \lambda (\xi ,1_n)\,\delta g=
{\textstyle\frac{1}{8}}\delta g\left( \frac{\pa\xi }{\pa x}\,+\,\frac{\pa\xi
}{\pa x}^T\right)  -{\textstyle\frac{1}{8}} \left( \frac{\pa\xi }{\pa
x}\,+\,\frac{\pa\xi }{\pa x}^T\right)\delta g.
\eee 
In other words we know
$$
\lambda (\xi ,1_n+h)= {\textstyle\frac{1}{2}} \left( \frac{\pa\xi }{\pa
x}\,-\,\frac{\pa\xi }{\pa x}^T\right)+ {\textstyle\frac{1}{8}}h\left(
\frac{\pa\xi }{\pa x}\,+\,\frac{\pa\xi }{\pa x}^T\right) 
=-{\textstyle\frac{1}{8}} \left( \frac{\pa\xi }{\pa x}\,+\,\frac{\pa\xi }{\pa
x}^T\right)h+O(h^2).
$$ To get the terms of order $h^2$, we use again  equation (\ref{commut})
 with $g=1_n+h=1_n+\delta g_2$ and compute the second functional derivative,
\bb
\frac{\delta^2 }{\delta g_1\delta g_2} \lambda (\xi ,1_n)\,\delta g_1\delta g_2:=
\frac{\delta }{\delta g_1} \lambda (\xi ,1_n+\delta g_2)\,\delta g_1-
\frac{\delta }{\delta g_1} \lambda (\xi ,1_n)\,\delta g_1+O(\delta g_2^2).
\eee 
Therefore
\bb
\lambda (\xi ,1_n+h)&=& {\textstyle\frac{1}{2}} \left(
\,\frac{\pa\xi }{\pa x}\,-\,\frac{\pa\xi }{\pa x}^T\,\right)+
{\textstyle\frac{1}{8}}h\left( \,\frac{\pa\xi }{\pa x}\,+\,\frac{\pa\xi }{\pa
x}^T\,\right)  -{\textstyle\frac{1}{8}} \left( \,\frac{\pa\xi }{\pa
x}\,+\,\frac{\pa\xi }{\pa x}^T\,\right)h\cr \cr &&
-{\textstyle\frac{1}{16}}h^2\left( \,\frac{\pa\xi }{\pa x}\,+\,\frac{\pa\xi
}{\pa x}^T\,\right)  +{\textstyle\frac{1}{16}} \left( \,\frac{\pa\xi }{\pa
x}\,+\,\frac{\pa\xi }{\pa x}^T\,\right)h^2 +O(h^3),
\eee 
reproducing the beginning of the Taylor series (\ref{series}). We also see
that  equation (\ref{commut}) fixes the terms of order
$h^{N+1}$ in
$\lambda (\xi ,1_n+h)$ from the terms of order $h^N$, which concludes the
proof.
\end{proof}

\begin{rem}

- As pointed out in \S 2, the trivial solution only makes sense on parallelizable
manifolds and, although used in the physics literature, we must discard it. So, 
in this restrictive sense, the Lie pseudoalgebra homomorphism is unique.

- Note that all terms of the series are local in the sense that they
contain no derivatives of $g$ or equivalently of $h$ with respect to $x$.
\end{rem}

Of course a proof for finite diffeomorphisms, and without resorting
to power series remains desirable.

\section{Lift and noncommutative geometry}

In this concluding section we would like to explain our motivation concerning
the uniqueness of the spin lift and an interesting, open problem it leads to.

We note that the spin lift from general relativity can be further extended
to noncommutative geometry
\cite{book,real,grav} where it then allows to define the configuration space of
the spectral action \cite{cc} via the fluctuations of the Dirac operator. For
inner automorphisms, these fluctuations have a natural motivation from
Morita equivalence, but they continue to make sense for outer automorphisms
like diffeomorphisms of a Riemannian spin manifold. In this commutative
case, the spectral action reproduces general relativity with a positive
cosmological constant plus a curvature square term. For almost commutative
geometries, the spectral action produces, in addition, the
complete Yang-Mills-Higgs action e.g. the standard model of electromagnetic,
weak and strong forces
\cite{cc}.

The commutative case relies on three steps: 
\begin{itemize}\item
Connes' reconstruction theorem \cite{grav} that motivates the definition of
spectral triples.
\item
A result on the uniqueness of the extension of the spin lift to all
diffeomorphisms.
\item
A result that puts Einstein's equivalence principle on a mathematical footing
by the use of Connes' fluctuating metric.
\end{itemize}
The guiding example to noncommutative geometry is any $n$-dimensional,
compact Riemannian spin manifold $M$. It defines a real spectral triple
$(\aaa,\hh,\dd)$ with algebra
$\aaa=\ccc^\infty(M)$ faithfully represented on the Hilbert space $\hh$ of
square integrable Dirac spinors and the selfadjoint Dirac operator
$\dd=\ddd$ possibly with torsion. The real structure is defined by the
anti-unitary operator
$J$ that physicists call charge conjugation when $M$ is interpreted as
space-time. For even dimensional manifolds $M$ there is another operator
$\chi $ defining a
$\zz_2$-grading of the Hilbert space. In physics it is called chirality, $\chi
=\gamma _5$. 

Some of the properties of the four or five items $\aaa,\ \hh,\  \dd,\ J,\ (\chi)$
are promoted to axioms of the (even), real spectral triple. The commutativity of
the algebra $\aaa=\ccc^\infty(M)$ is not promoted and the triple is called
commutative if its algebra is. The selection of the promoted properties is
motivated by  Connes' reconstruction theorem
\cite{grav}. It states that every commutative, (even), real spectral triple comes
from an (even dimensional) Riemannian spin manifold. An explicit proof of a
weaker form of the reconstruction theorem can be found in the Costa Rica book
\cite[Theorem 11.2]{costa}.

If $M$ is interpreted as phase space, then $\aaa$ is the algebra of classical
observables. Not promoting its commutativity opens the door to Heisenberg's
uncertainty relation.

In general relativity, Einstein generalizes rotations to general coordinate
transformations. Let us momentarily ignore coordinate singularities and
confuse coordinate transformation and diffeomorphism. In the language of
spectral triples, diffeomorphisms are algebra automorphisms,
Aut$\big(\ccc^\infty(M)\big)=$ Diff$(M)$. Our aim is therefore to lift any
algebra automorphism $\sigma \in {\rm Aut}(\aaa)$ to unitaries on the
Hilbert space $\hh$: $\sigma \mapsto \LL(\sigma ).$ In addition to being
unitary  we want the lifted automorphism to commute with the real structure
and in the even case with the chirality, and to satisfy the covariance property
\bb
 \LL(\sigma )\rho (a) \LL(\sigma )^{-1}=\rho \big(\sigma (a)\big),\qq {\rm
for\ all}\ a\in \aaa,
\ee where we denote by $\rho $ the faithful representation of $\aaa$ on
$\hh$.  This property allows to define the projection of the lift back to the
automorphism group
\bb p\big(\LL(\sigma )\big)(a)=\rho ^{-1}\big(\LL(\sigma )\rho
(a)\LL(\sigma )^{-1}\big).
\ee
 Of course we also want
$\LL$ to be a group homomorphism, possibly multi-valued.

In the commutative case, if $U$ is a coordinate neighborhood in $M$ we have
for diffeomorphisms close to the identity
\bb \LL\big({\rm Diff}(U)\big)={\rm Diff}(U)\ltimes \,^U Spin(n).\ee
 In coordinates a possible lift is
$\LL=(1,S\circ
\Lambda) $ where $S$ is the usual double valued homomorphism (\ref{spin}).

\subsection{Equivalence principle}

Einstein uses the equivalence principle to guess the configuration space of
general relativity: His starting point is the flat metric in inertial coordinates.
With a general coordinate transformation he arrives at the metric tensor
relevant for a uniformly accelerated observer. For this observer, a free,
massive point-particle is subject to a constant pseudo-force, which looks like
gravity on earth. Therefore he proposes nontrivial metric tensors to encode
gravity. If we want to formulate Einstein's idea in mathematical terms, we face
the problem that a flat metric will still be flat after a general coordinate
transformation and to express the equivalence between  the free particle as
seen by the accelerated observer and the particle  falling freely in a static
gravitational field as seen by an observer at rest with respect to the
gravitational field we need more than one coordinate transformation. 

This situation is somehow reminiscent of special relativity and
electromagnetism. The magnetic field of a constant rectilinear current is a
pseudo-force, which can be transformed to zero with the boost  that puts the
observer at rest with respect to the charges. A  general magnetic field could be
viewed as generated by a superposition of many small rectilinear currents and
one is tempted to use many `local' boosts to transform it to zero.

In a sense this is what Alain Connes does when he {\it fluctuates} the metric.
In his reconstruction theorem, the metric is reconstructed from and therefore
encoded in the Dirac operator. Now Connes replaces the point-like matter in
Einstein's reasoning by a Dirac particle. Then the Dirac operator $\dd$ plays
two roles simultaneously, (i) it defines the metric and therefore the initial, say
zero, gravitational field,  (ii) it defines the dynamics of matter. Using the spin
lift $\LL(\sigma )$ we can act with a general coordinate transformation on the
Dirac particle
$\psi
\in \hh$. In a coordinate neighborhood, this action reads
\bb \big( \LL(\sigma  )\psi \big) ( x)=\big(S\left(\Lambda (\sigma,g
)\big)\right|_{\sigma  ^{-1}( x)}\psi
\big({\sigma ^{-1}( x)}\big),\ee and amounts to replacing the initial flat Dirac
operator $\dd$ by the still flat operator $\LL(\sigma )\,\dd\,\LL(\sigma)
^{-1}$. So far we have not gained anything with respect to Einstein's point of
view. However, unlike metrics, Dirac operators can be linearly combined and
Connes defines the fluctuations of the initial metric or Dirac operator as the
finite linear combinations 
\bb \sum r_j\,\LL(\sigma _j)\,\dd\,\LL(\sigma _j)^{-1}, \qq r_j\in\rr,\qq
\sigma _j\in{\rm Aut}(\aaa).\ee
 We arrive naturally at the following
\begin{pb} Given two Dirac operators  with arbitrary curvature and torsion
defined on the same manifold, can one be written as a fluctuation of the other?
\end{pb} Of course, the configuration space of general relativity, the space of
all gravitational fields, should be identified as the {\it affine} space of all
fluctuations of the initial Dirac operator, from which however we delete those
fluctuations that do not define a spectral triple, e.g. the fluctuation that is
identically zero. 

For almost commutative spectral triples this affine space in addition contains
the Yang-Mills connections and the Higgs scalar \cite{real, grav}.

\vskip.5 cm \noindent {\bf Acknowledgement:} It is a pleasure to thank Antony
Wassermann for help and advice.

\end{document}